# Optimizing Multi-UAV Deployment in 3D Space to Minimize Task Completion Time in UAV-Enabled Mobile Edge Computing Systems

Sujunjie Sun, Guopeng Zhang, Haibo Mei, Kezhi Wang, and Kun Yang

*Abstract*—In Unmanned Aerial Vehicle (UAV)-enabled mobile edge computing (MEC) systems, UAVs can carry edge servers to help ground user equipment (UEs) offloading their computing tasks to the UAVs for execution. This paper aims to minimize the total time required for the UAVs to complete the offloaded tasks, while optimizing the three-dimensional (3D) deployment of UAVs, including their flying height and horizontal positions. Although the formulated optimization is a mixed integer nonlinear programmming, we convert it to a convex problem and develop a successive convex approximation (SCA) based algorithm to effectively solve it. The simulation results show that the joint optimization of the horizontal and the vertical position of a group of UAVs can achieve better performance than the traditional algorithms.

*Index Terms* — UAV, MEC, 3D deployment, successive convex approximation.

## I. INTRODUCTION

With the development of information and communication technology (ICT), smart home, wearable devices and Internet of things (IoTs) are becoming more and more popular in our daily life. An increasing number of computationally-expensive and delay-sensitive mobile applications, such as virtual reality, augmented reality, object detection, etc., make them difficult to be executed by mobile user equipment (UEs) due to their limited computing and energy resources. Fortunately, mobile edge computing (MEC) [1] provides an effective solution to the above problem by deploying computing resources into network edge (e.g., cellular base stations or BSs). In this case, UEs can offload their computing tasks to the edge servers for execution. However, in some special cases, the network infrastructure could be damaged by natural disasters or the computing and communication resources at the edge servers may not be adequate due to rush hours, alternatives are expected to be deployed.

Recently, UAV has been developed rapidly, thus providing the solution to the problem mentioned above. UAVs may carry edge servers to provide the computing services on demand [2]. Additionally, there is high possibility for the UAV to establish the line of sight (LoS) to the ground UEs. Although the computing resources carried by UAVs are limited compared with ground edge servers, multiple UAVs can be applied to add both flexibility and capacity.

Sujunjie Sun and Guopeng Zhang are both with the School of Computer Science and Technology, China University of Mining and Technology, Xuzhou, China, 221116. (email: sujunjie_sun@cumt.edu.cn, gpzhang@cumt.edu.cn)

Haibo Mei is with the School of Communication and Information Engineering, University of Electronic Science and Technology of China, Chengdu, China, 611731. (email: haibo.mei@uestc.edu.cn)

Kezhi Wang is with the Department of Computer and Information Science, Northumbria University, Newcastle NE2 1XE, UK. (email: kezhi.wang@northumbria.ac.uk)

Kun Yang is with the School of Computer Science and Electronic Engineering, University of Essex, Wivenhoe Park, Colchester, CO4 3SQ, UK. (email: kunyang@essex.ac.uk)

This work was supported by the National Natural Science Foundation of China (Grant nos. 61971421)

In [3], the authors used the UAV to wirelessly charge the ground IoT devices and provide computing services. By jointly optimizing the hovering time of the UAV and the services sequence of the IoT devices, the overall energy consumption of the UAV can be minimized. The authors of [4] minimized the energy consumption of multiple UAVs by optimizing the UAV deployment. But they only optimized the horizontal location of the UAVs, assuming their heights were fixed. To balance the computing load among multiple UAVs, the authors of [5] proposed a differential-evolution based UAV deployment strategy and a deep reinforcement learning based algorithm to schedule the tasks for each UAV. In addition to optimizing the energy consumption of UAVs, the authors in [6] and [7] also proposed to optimize the total time consumed by UAVs to complete the computing or communication tasks of ground UEs.

Inspired by the above works, this paper studies the 3D spatial deployment of multiple UAVs in UAV-enabled MEC systems. The main contributions include the following two aspects:

1) Multiple UAVs not only provides computation resource to ground UEs, but also constitutes a parallel computing system. In order to achieve the objective, i.e., minimizing the time required to complete all the computing tasks offloaded by the UEs, we formulate the UAV deployment problem as a *min-max* problem to minimize the operating time of the UAV which takes the longest time to complete the tasks among all UAVs.

2) We adopt the Rician fading channel model [8] as it describes the fading property of the UAV-to-UE link in 3D space more accurately. This not only introduces several non-convex constraints for the *min-max* problem, but also needs to jointly optimize the vertical and horizontal positions of UAVs and the association between UAVs and ground UEs. This paper presents an efficient algorithm to obtain the approximate solution to this *Mixed Integer Nonlinear Programming* (MINP) problem.

The remainder of this paper is organized as follows. In Section II, we introduce the system model and problem formulation. In Section III, the proposed solution is given. Simulation is given in Section IV, whereas conclusion is provided in Section V.

## II. SYSTEM MODEL

We consider a typical scenario of UAV-enabled MEC system, as shown in Fig. 1. Let $N$ denote the number of UEs, and $M$ as the number of UAVs dispatched. The set of the UEs and the UAVs are represented by $\mathcal{N} = \{1,2,\ldots,N\}$ and $\mathcal{M} = \{1,2,\ldots,M\}$, respectively. We consider a 3D Euclidean coordinate system. The position of the $i$th ($\forall i \in \mathcal{N}$) UE is represented by $(\boldsymbol{w}_i, 0)$, with $\boldsymbol{w}_i = (x_i, y_i) \in \mathbb{R}^{1\times 2}$. The position of $j$th UAV is represented by $(\boldsymbol{q}_j, H_j)$, with $\boldsymbol{q}_j = (X_i, Y_i) \in \mathbb{R}^{1\times 2}$. It should be noted that the position of the UEs is a priori, but the 3D coordinate $(X_j, Y_j, H_j)$ of each UAV is to be optimized.

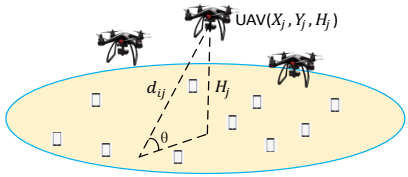

Fig. 1 A UAV-enabled MEC system

We assume that the $i$th UE has a computing task $I_i$ to be executed which is characterized by a two-tuple of parameters $(D_i, F_i)$, where $D_i$ (in bits) represents the amount of the input data, and $F_i$ (in CPU cycles) represents the total number of CPU cycles of required by the task. The UEs offload their computing tasks to the UAVs through wireless transmissions. We assume that a set of $M$ non-overlapping channel are assigned to the UAVs for collecting the task input data from the UEs, and the UEs associated with the same UAV upload their data using TDMA technology without causing co-channel interference. It is reasonable to assume that all UEs expect to upload their task input data as soon as possible. So we set the transmission power of each UE, i.e., $p_i$, to be constant (e.g., the maximum allowable transmission power). Let $r_{i,j}$ denote the fixed transmission rate from the $i$th UE to the $j$th UAV. Then, the expected time to upload the task input data is given by $t_{i,j}^U = D_i / r_{i,j}$.

In actual wireless transmission, the channel between UAVs and UEs are stochastic fading, and the channel coefficient from the $i$th UE to the $j$th UAV is modelled as
$$h_{i,j} = \sqrt{\beta_{i,j}} g_{i,j}. \quad (1)$$
where $\beta_{i,j}$ is the large-scale average channel power gain and $g_{i,j}$ is the small-scale fading coefficient.

Let $d_{i,j} = \sqrt{\|q_j - w_i\|^2 + H_j^2}$ denote the distance between the $i$th UE and the $j$th UAV. The distance-dependent average channel power gain $\beta_{i,j}$ is given by
$$\beta_{i,j} = \beta_0 (d_{i,j})^{-a_0}. \quad (2)$$
where $a_0$ is the path-loss exponent and $\beta_0$ is the channel power gain at a reference distance of $d = 1$ m.

Due to the existence of LoS components in UAV-UE links, the small-scale fading $g_{i,j}$ is modelled as Rician fading as
$$g_{i,j} = \sqrt{\frac{k_{i,j}}{k_{i,j}+1}} g + \sqrt{\frac{1}{k_{i,j}+1}} \tilde{g}. \quad (3)$$
where $g$ corresponds to the LoS component with $|g| = 1$, $\tilde{g}$ corresponds to the Rayleigh fading component which is a zero-mean unit-variance circularly symmetric complex Gaussian random variable, and $k_{i,j}$ is the Rician factor which is the ratio of the power in the LoS component to that in the fading component. Let $\theta_{i,j} = \arcsin(H_j / d_{i,j})$ denote the elevation angle between a UE and a UAV. The Rician factor $k_{i,j}$ is closely related to $\theta_{i,j}$, which is given by
$$k_{i,j} = A_1 \exp(A_2 \theta_{i,j}). \quad (4)$$
where $A_1$ and $A_2$ are constants determined by the environment.

Therefore, due to the transmission of the $i$th UE, the maximum achievable rate at the $j$th UAV can be given by
$$C_{i,j} = B \log_2 \left(1 + \frac{|h_{i,j}|^2 p_i}{\sigma^2 \Gamma}\right). \quad (5)$$
where $\sigma^2$ is the noise power at the receiver of the UAV, $B$ is the channel bandwidth, and $\Gamma > 1$ is the signal-to-noise ratio (SNR) gap. Once the $i$th UE decides to offload the task $I_i$ to the $j$th UAV, the actual minimum time to upload the task input data is given by $\widetilde{t_{i,j}^U} = D_i / C_{i,j}$.

Let $P_{i,j}$ denote that the timeout probability that the $j$th UAV cannot successfully receive the data from the $i$th UE within time $t_{i,j}^U$. We can denote $P_{i,j}$ by
$$P_{i,j} = \mathbb{P}\left(\widetilde{t_{i,j}^U} > t_{i,j}^U\right) = F_{i,j}\left(\frac{\sigma^2 \Gamma (2^{r_{i,j}} - 1)}{\beta_{i,j} p_i}\right). \quad (6)$$
where $F_{i,j}$ is the non-decreasing cumulative distribution function (CDF) of the random variable $|g_{i,j}|^2$ w.r.t $r_{i,j}$.

In order to make the timeout probability lower than a certain level $\varepsilon_0$, we set $P_{i,j} = \varepsilon_0$ and $0 < \varepsilon_0 \leq 0.1$ for $\forall i,j$. By substituting eq. (3) into (6) with $P_{i,j} = \varepsilon_0$, we get
$$r_{i,j} = B \log_2 \left(1 + \frac{\varphi_{i,j} \gamma_i}{\left((X_j - x_i)^2 + (Y_j - y_i)^2 + H_j^2\right)^{a_0/2}}\right) \quad (7)$$
where $\gamma_i \triangleq \frac{p_i \beta_0}{\sigma^2 \Gamma}$ and $\varphi_{i,j}$ denotes the solution to $F_{i,j}(x) = \varepsilon_0$. Because it is difficult to get a closed-form expression of $\varphi_{i,j}$, we used the method provided by ref. [8] and approximate $\varphi_{i,j}$ by the logistic function. Finally, we have
$$r_{i,j} \approx B \log_2 \left(1 + \left(K_1 + \frac{K_2}{1 + \exp(-(K_3 + K_4 v_{i,j}))}\right) \frac{\gamma_i}{(d_{i,j})^{a_0}}\right) \quad (8)$$
where $K_3 < 0$ reflects the positive logistic mid-point, $K_4 > 0$ is the logistic growth rate, $K_1 > 0$ and $K_2 > 0$ satisfy $K_1 + K_2 = 1$, and $v_{i,j} \triangleq \sin(\theta_{i,j})$.

### III. PROBLEM FORMULATION

Based on the system model described above, the specific scheme of task offloading and execution at any $j$th UAV is shown in Fig. 2. The structure of the service flow consists of two phases, i.e., the UE task offloading phase and the UAV computation phase. The computing phase begins after the end of the task offloading phase, and these two phases cannot be executed in parallel.

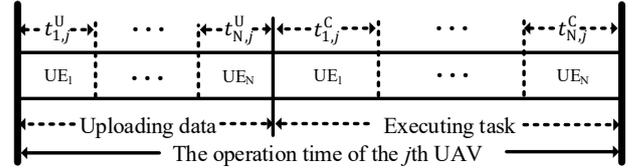

Fig. 2 Computing task offloading and execution at the $j$th UAV

Without loss of generality, we assume that the $i$th UE offloads the computing task $I_i$ as a whole to any $j$th UAV for remote execution. Let $\alpha_{i,j} \in \{0,1\}$ denote the offloading decision of the $i$th UE. If $\alpha_{i,j} = 1$, it means that the $i$th UE decides to offload the task to the $j$th UAV, $\alpha_{i,j} = 0$, otherwise. We also assume that the $i$th UE can only select one UAV to offload its task, while any $j$th UAV can serve multiple UE during the operation. Hence, we have the constraints:
$$\sum_{j=1}^{M} \alpha_{i,j} = 1, \alpha_{i,j} \in \{0,1\}, \forall i \in \mathcal{N}, \forall j \in \mathcal{M} \quad (9)$$
Let $f_{i,j}$ denote the computing resource (in CPU cycles per second) assigned to the $i$th UE by the $j$th UAV. Then, the time required to perform the computing task $I_i$ at the $j$th UAV is
$$t_{i,j}^C = \frac{F_i}{f_{i,j}}, \forall i \in \mathcal{N}, \forall j \in \mathcal{M} \quad (10)$$
Let $f_j^{\max}$ denote the maximum computing power of the $j$th UAV. Then, the constraints of computing resource are
$$0 \leq f_{i,j} \leq f_j^{\max}, \forall i \in \mathcal{N}, \forall j \in \mathcal{M} \quad (11)$$
We aim to minimize the time required to complete all the computing tasks offloaded by the UEs. As the multiple UAV system constitutes a parallel computing paradigm, this can be achieved by minimizing the operating time of the UAV which takes the longest time to complete the tasks among all UAVs. This optimization problem is formulated as

$$\min_{\alpha_{i,j}, q_j, H_j, v_{i,j}} \max_{\forall j \in \mathcal{M}} \sum_{i=1}^{N} \alpha_{i,j}\left(t_{i,j}^{\text{U}}(q_j, H_j, v_{i,j}) + t_{i,j}^{\text{C}}\right) \quad (12)$$

$$\text{s.t.} \quad \alpha_{i,j} \in \{0,1\}, \forall i \in \mathcal{N}, \forall j \in \mathcal{M} \quad (12.1)$$

$$\sum_{j=1}^{M} \alpha_{i,j} = 1, \forall i \in \mathcal{N} \quad (12.2)$$

$$v_{i,j} = \frac{H_j}{\sqrt{\|q_j - w_i\|^2 + H_j^2}} \quad (12.3)$$

$$0 \leq f_{i,j} \leq f_j^{\max}, \forall i \in \mathcal{N}, \forall j \in \mathcal{M} \quad (12.4)$$

$$X^{\min} \leq X_j \leq X^{\max}, \ Y^{\min} \leq Y_j \leq Y^{\max} \quad (12.5)$$

$$H^{\min} \leq H_j \leq H^{\max} \quad (12.6)$$

In the problem, constraints (12.5) and (12.6) are used to limit the range of the horizontal and vertical motion of a UAV, respectively. $X^{\min}$, $Y^{\min}$ and $H^{\min}$ are the allowable minimum x, y and z axes coordinates for the UAV, and $X^{\max}$, $Y^{\max}$ and $H^{\max}$ are the maximum x, y and z axes coordinates of a UAV, respectively. It is noted that the objective function of problem (12) is a non-convex function w.r.t variables $\alpha_{i,j}, q_j, H_j$, and $v_{i,j}$, which makes the joint optimization problem challenging to solve. Additionally, constraint (12.3) is non-convex which is also difficult to deal with. Therefore, there is no standard way to solve this Mixed Integer Nonlinear Programming (MINP) problem directly.

## IV. Problem Solution

In this section, we present an effective method to find the approximate solution of problem (12). First, we define

$$\mu \triangleq \max_{\forall j \in \mathcal{M}} \sum_{i=1}^{N} \alpha_{i,j}(t_{i,j}^{\text{U}} + t_{i,j}^{\text{C}}) \quad (13)$$

as the operation time of the UAV which takes the longest time to complete the tasks among all the UAVs. Then, problem (12) can be rewritten as

$$\min_{\alpha_{i,j}, q_j, H_j, v_{i,j}, \mu} \mu \quad (14)$$

$$\text{s.t.} \quad (12.1) \sim (12.6) \quad (14.1)$$

$$\mu \geq \sum_{i=1}^{N} \alpha_{i,j}(t_{i,j}^{\text{U}} + t_{i,j}^{\text{C}}), \forall j \in \mathcal{M} \quad (14.2)$$

By relaxing the equality constraint (12.3) of problem (14) to an inequality constraint, we get the following problem.

$$\min_{\alpha_{i,j}, q_j, H_j, v_{i,j}, \mu} \mu \quad (15)$$

$$\text{s.t.} \quad (12.1),(12.2),(12.4) \sim (12.6),(14.2) \quad (15.1)$$

$$v_{i,j} \leq \frac{H_j}{\sqrt{\|q_j - w_i\|^2 + H_j^2}}, \forall i \in \mathcal{N}, \forall j \in \mathcal{M} \quad (15.2)$$

**Theorem 1:** The solution of problem (15) is equivalent to the solution of problem (14).

*Proof*: Please refer to Appendix A.

In order to solve problem (15), our method is to decompose the problem into three subproblems. Then, the approximate solution of problem (15) can be obtained by solving the three subproblems iteratively. The goal of the first subproblem is to find the optimal UAV and UE association $\alpha_{i,j}$ for a given UAV deployment $\{q_j, H_j\}$. Then, in the second subproblem, the aim is to find the optimal UAV horizontal position $q_j$ with the known $\alpha_{i,j}$ and $H_j$. Finally, the aim of the third subproblem is to find the optimal UAV vertical position $H_j$ with the known $\alpha_{i,j}$ and $q_j$.

### A. Association Optimization

For a given UAV deployment, i.e., $q_j$ and $H_j$ for $\forall j$, we relax the binary variables $\alpha_{i,j} \in \{0,1\}$ in constraint (12.1) into the continuous one as $0 \leq \widehat{\alpha_{i,j}} \leq 1$, and can get the following problem.

$$\min_{\alpha_{i,j}, \mu} \mu \quad (16)$$

$$\text{s.t.} \quad (12.2), (12.4), (14.2) \quad (16.1)$$

$$0 \leq \widehat{\alpha_{i,j}} \leq 1, \forall i \in \mathcal{N}, \forall j \in \mathcal{M} \quad (16.2)$$

Since all the constraints of problem (16) are linear, this problem is a linear programming (LP) problem, which can be solved by using standard solution method. As for some $\widehat{\alpha_{i,j}}$ take the values between 0 and 1, we adopt the following recovery method as in [2] to force their values to be 0 or 1.

$$\alpha_{i,j} = \begin{cases} 1, & \widehat{\alpha_{i,j}} \geq 0.5 \\ 0, & \widehat{\alpha_{i,j}} < 0.5 \end{cases} \quad (17)$$

It is noted that there exists only one $\widehat{\alpha_{i,j}}$ greater than 0 on any row of the association matrix as $0 \leq \widehat{\alpha_{i,j}} \leq 1$. This recovery method ensures that the association matrix does not have multiple elements with a value of 1 in each row. Therefore, the obtained $\alpha_{i,j}$ is a feasible solution to the original problem (12). It should be also noted that this solution method can only obtain the approximate optimal solution of this type of 0-1 programming problems. However, this method is with lower computational complexity, especially when the numbers of the UAVs and the UEs are large (in such a situation, the optimal solution of this problem cannot be obtained by using regular method, e.g., the complete enumeration method).

### B. Horizontal Position Optimization

For a given UAV vertical position $H_j$ and the obtained UAV and UE association $\alpha_{i,j}$, one can get the problem of optimizing the UAV horizontal position as

$$\min_{q_j, v_{i,j}, \mu} \mu \quad (18)$$

$$\text{s.t.} \quad (12.4), (12.5), (15.2), (14.2) \quad (18.1)$$

It is noted that constraints (12.4) and (12.5) are linear, and constraints (15.2) and (14.2) are non-convex. Hence, problem (18) is a non-convex optimization problem. To tackle this difficulty, we define a slack variable $z_{i,j}$ for $\forall i,j$, and rewrite problem (18) as

$$\min_{q_j, v_{i,j}, z_{i,j}, \mu} \mu \quad (19)$$

$$\text{s.t.} \quad (12.4), (12.5), (15.2) \quad (19.1)$$

$$\mu \geq \sum_{i=1}^{N} \alpha_{i,j}\left(\frac{D_i}{z_{i,j}} + \frac{F_i}{f_{i,j}}\right), \forall j \in \mathcal{M} \quad (19.2)$$

$$z_{i,j} \leq r_{i,j} \quad (19.3)$$

**Theorem 2:** The solution of problem (19) is equivalent to the solution of problem (18).

*Proof*: Please refer to Appendix B.

In order to solve problem (19), we need to apply the successive convex approximation (SCA) method. Then we can convert constraints (19.3) and (15.2) into convex constraints by the following two theorems.

**Theorem 3:** For a given $\widehat{q_j}$, there exists a definite lower bound $r_{i,j}^{\text{lb}}$ for $r_{i,j}$, which is given by

$$r_{i,j} \geq r_{i,j}^{\text{lb}} \triangleq \widehat{r_{i,j}} - \psi_{i,j}^{\text{X}}\left(e^{-(K_3 + K_4 v_{i,j})} - e^{-(K_3 + K_4 \widehat{v_{i,j}})}\right)$$
$$- \psi_{i,j}^{\text{Y}}\left(\|q_j - w_i\|^2 - \|\widehat{q_j} - w_i\|^2\right) \quad (20)$$

The definitions of $\psi_{i,j}^{\text{X}}$ and $\psi_{i,j}^{\text{Y}}$ are shown at the top of the next page.

*Proof:* Please refer to Appendix C.

**Theorem 4:** For a given $\widehat{q_j}$, there exists a definite lower bound $v_{i,j}^{\text{lb}}$ for $v_{i,j}$, which is given by

$$\psi_{i,j}^{\text{X}} = \frac{\gamma A_2 B}{\ln 2 \left(1+e^{-(K_3+K_4\widehat{v_{i,j}})}\right)\left(\gamma\left(A_1\left(1+e^{-(K_3+K_4\widehat{v_{i,j}})}\right)+A_2\right)+\left(1+e^{-(K_3+K_4\widehat{v_{i,j}})}\right)\left(\widehat{d_{i,j}}^2\right)^{a_0/2}\right)}$$

$$\psi_{i,j}^{\text{Y}} = \frac{\gamma a_0 B\left(A_1\left(1+e^{-(K_3+K_4\widehat{v_{i,j}})}\right)+A_2\right)}{\ln 4 \widehat{d_{i,j}}^2 \left(\gamma\left(A_1\left(1+e^{-(K_3+K_4\widehat{v_{i,j}})}\right)+A_2\right)+\left(1+e^{-(K_3+K_4\widehat{v_{i,j}})}\right)\left(\widehat{d_{i,j}}^2\right)^{a_0/2}\right)} \quad (21)$$

$$v_{i,j}^{\text{lb}} \triangleq \widehat{v_{i,j}} - \frac{H_j}{2\left(\|\widehat{q_j}-w_i\|^2+H_j^2\right)^{3/2}} \times \left(\|q_j-w_i\|^2 - \|\widehat{q_j}-w_i\|^2\right) \quad (22)$$

*Proof:* Please refer to Appendix D.

Using **Theorems 3 and 4**, we can convert problem (19) into the following convex problem.

$$\min_{q_j, v_{i,j}, z_{i,j}, \mu} \mu \quad (23)$$

$$\text{s.t.} \quad (12.4),(12.5),(19.2) \quad (23.1)$$
$$v_{i,j} \leq v_{i,j}^{\text{lb}} \quad (23.2)$$
$$z_{i,j} \leq r_{i,j}^{\text{lb}} \quad (23.3)$$

This problem can be solved effectively by using some optimization toolbox, e.g., CVX.

### C. Vertical Position Optimization

For a given UAV horizontal position $q_j$ and UAV and UE association $\alpha_{i,j}$, the UAV vertical position can be optimized by solving the following problem

$$\min_{H_j, v_{i,j}, z_{i,j}, \mu} \mu \quad (24)$$
$$\text{s.t.} \quad (12.4), (12.6), (15.2), (19.2), \text{ and } (19.3) \quad (24.1)$$

It is easy to know that constraints (12.4) and (12.6) are linear, and constraint (19.2) is convex. The following theorem shows that constraint (15.2) is also convex.

**Theorem 5:** Constraint (15.2) is convex with respect to $v_{i,j}$ and $H_j$.

*Proof:* Please refer to Appendix E.

However, constraint (19.3) is non-convex. Next, we apply the SCA method and convert it into a convex one by the following theorem.

**Theorem 6:** For a given $\widehat{H_j}$, there exists a definite lower bound $r_{i,j}^{\text{lb}}$ for $r_{i,j}$, which is given by

$$r_{i,j} \geq \widetilde{r_{i,j}^{\text{lb}}} \triangleq \widehat{r_{i,j}} - \phi_{i,j}^{\text{X}} \left(e^{-(K_3+K_4 v_{i,j})} - e^{-(K_3+K_4\widehat{v_{i,j}})}\right) - \phi_{i,j}^{\text{Y}}(H_j^2 - \widehat{H_j}^2) \quad (25)$$

The coefficients $\phi_{i,j}^{\text{X}}$ and $\phi_{i,j}^{\text{Y}}$ are given in Appendix F

*Proof:* Please refer to Appendix F.

Using **Theorem 6**, we can convert problem (24) into the following convex problem.

$$\min_{H_j, v_{i,j}, z_{i,j}, \mu} \mu \quad (26)$$
$$\text{s.t.} \quad (12.4),(12.6),(15.2),(19.2) \quad (26.1)$$
$$z_{i,j} \leq \widetilde{r_{i,j}^{\text{lb}}} \quad (26.2)$$

It can be solved efficiently by using some optimization toolbox, e.g., CVX.

### D. Overall Algorithm Description

In this section, we present the overall algorithm for addressing problem (12) in the following **Algorithm 1**.

In Algorithm 1, the selection of different initial points will influence the convergence speed of the algorithm. In order to ensure the generality of the algorithm, we do not rely on sophisticated methods to set the initial point. In the following simulations, we find that the algorithm converges to the stable point after about 20 iterations by randomly selecting a point as the initial position within the feasible domain satisfying the problem constraints.

---

**Algorithm 1** Overall algorithm to solve problem (12)

1. **Initialize:** Set the iteration number $r = 0$.
   Find a feasible UAV deployment $\{q_j, H_j\}$.
2. **Repeat:**
3. For a given UAV deployment $\{q_j^r, H_j^r\}$, obtain the optimal UAV and UE association, denoted by $a_{i,j}^{r+1}$, by solving the LP problem (16).
4. Based on the obtained $a_{i,j}^{r+1}$ and $H_j^r$, use *CVX toolbox* to obtain the solution to problem (23) denoted as $q_j^{r+1}$.
5. Based on the obtained $\{q_j^{r+1}, a_{i,j}^{r+1}\}$, use *CVX toolbox* to obtain the solution to problem (26) denoted as $H_j^{r+1}$.
6. Update $r = r + 1$.
7. **Until:** The objective value $\mu$ of problem (12) converges, or the maximum number of iterations ($r^{\max}$) is reached.
8. **Return:** The UAV deployment $\{q_j^*, H_j^*\}$, and the UAV and UE association $a_{i,j}^*$.

---

## V. SIMULATION RESULTS

This section shows the simulation results to testify the performance of the proposed method. We consider a 100×100 m² rectangular area, where the UEs are randomly distributed on the ground. For each UAV, the minimum and the maximum heights are set to be $H^{\min} = 40$ m, $H^{\max} = 80$ m, respectively. The bandwidth is $B = 10$ MHz. The other parameters used in the simulation are given below. The channel power gain at the reference distance of 1 m is $-30$ dB. The noise power at the receiver of a UAV is $10^{-9}$ W. The transmit power of each UE is 20 dBm, and the SNR gap $\Gamma = 8.2$ dB. The parameters of the logistic function are set as $K_1 = 0.01$, $K_2 = 0.99$, $K_3 = -4.7$, and $K_4 = 8.9$. The available computing resource of a UAV is 2 GB/s. The numbers of CPU cycles required to compute each data bit is 300 (in cycles/bit).

To testify the performance of the proposed method, we compare it with three other methods, termed as horizontal position optimization (HPO), vertical position optimization (VPO), and conventional LoS-link based optimization (CLBO). In the HPO method, only the horizontal positions of the UAVs are optimizing, while their heights are fixed at 60m. In the VPO, we first cluster the ground UEs and use *K*-means algorithm to find their horizontal central points. Then we optimize the height of the UAVs to minimize the mission completion time. In the CLBO, the UE and UAV association and the UAV 3D position are optimized by using the conventional LoS based channel model [9].

First, we show in Fig. 3 the optimal UAV deployment in the system and the association between the UAVs and the UEs. The numbers of the UAVs and the UEs are 3 and 30, respectively.

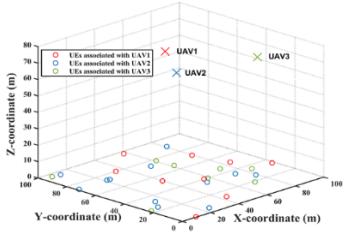

Fig. 3 3D UAV deployment and UE-to-UAV association.

Intuitively, when a UE is closer to a UAV, it may have better channel condition and higher data rate for uploading their task input data. However, the computation capacity of a UAV is limit, and it cannot process the tasks offloaded by all the UEs close to it. Fig. 3 shows that the presented algorithm can effectively distribute the computation loads of the ground UEs among the UAVs, as the association between the UAVs and the UEs is not entirely determined by their channel conditions.

Next, we employ five UAVs in the system and investigate the trend of the mission completion time with the increase of the number of the UEs from 10 to 80. The simulation result is shown in Fig. 4.

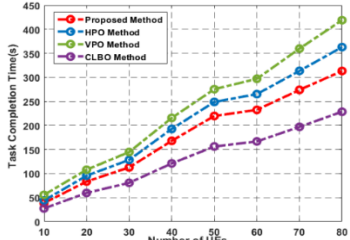

Fig. 4 The variation of task completion time with the number of UEs.

One can see that the mission completion time of all the methods increase as the number of UEs increases, and our proposed method outperforms the HPO and VPO methods in all situations. As the UE number increases, the performance gap between the proposed method and the HPO/VPO method increases gradually. It indicates that the joint optimization of the horizontal and vertical positions of the UAVs is better than optimizing the horizontal position only or the vertical position only. Moreover, we find that the performance loss of our method is small in comparison to the CLBO. It verifies the effectiveness of the proposed method. The reason is that the proposed method can not only utilize the LoS but also the fading characteristics of the UAV and UE links in 3D space. It thus leads to a more suitable 3D deployment of the UAVs to complete the UE offloaded tasks.

Next, we fix the number of the UEs to 80, and increase the number of the UAVs from 5 to 10. The simulation result is shown in Fig. 5.

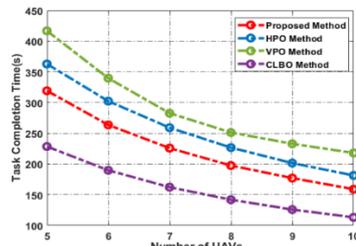

Fig. 5 The variation of task completion time with the number of UAVs.

From Fig. 5, one can see that the mission completion time of all the methods decreased with the increase of the number of UAVs. This is because more UAVs can bring more computing resources to UEs, and each UAV can associate less UE and provide better service. Moreover, for different numbers of UAVs, the proposed method is again superior to the VPO and HPO methods, which verifies that the proposed method can well balance the computation and communication loads of the UEs among the UAVs.

Finally, it is noted that the simulation platform is on a PC with Intel i7 2.90 GHz CPU and 16-GB memory. The software is MATLAB (2019b) plus CVX (V2.2). When we use 5 UAVs and 50 UEs, the execution time of the algorithm is 671 s. Considering practical UAV scheduling platform should be with more powerful servers, less computing time is required. The above simulation results indicate that joint optimization of the horizontal and vertical positions of a group of UAVs can achieve better performance than the single optimization of the horizontal position or the vertical position of them.

## VI. CONCLUSION

In this paper, we have optimized multi-UAV deployment to minimize the time required to complete the computing tasks offloaded by ground UEs. Although it is a MINLP problem, we have proposed an effective algorithm based on SCA to address it successfully. The performance of the proposed algorithm has been verified by simulation and considerable performance has been demonstrated.

## APPENDIX A

*Proof of Theorem 1*: It can be shown that the solution of problem (15) is equivalent to the solution of problem (14) when the equality of constraint (15.2) holds. Obviously, in the case of $v_{i,j} < \frac{H_j}{\sqrt{\|q_j - w_i\|^2 + H_j^2}}$, $v_{i,j}$ is negatively correlated with the objective value of Problem (15) while the other variable are fixed. As $v_{i,j}$ infinitely approximates with the equation value, the objective value of Problem (15) will further improved. Therefore, problem (15) is equivalent to problem (14). □

## APPENDIX B

*Proof of Theorem 2:* Obviously, problem (19) is equivalent to problem (18) when the equality of constraint (19.3) holds. In the case of $z_{i,j} \neq r_{i,j}$, then $\mu$ always remains constant with the increasing $z_{i,j}$ while the other variables are fixed. Therefore, problem (19) is still equivalent to problem (18). □

## APPENDIX C

*Proof of Theorem 3*: It can be showed that $f(x,y) = B\log_2\left(1 + \left(A_1 + \frac{A_2}{x+X}\right)\frac{\gamma}{(y+Y)^{a_0/2}}\right)$ is a convex function w.r.t x and y. Then, we can use the first-order Taylor expansion to approximate the convex function $f(x,y)$ at any $x_0$ and $y_0$. Therefore, we can get

$$f(x,y) \geq f(x_0, y_0) + f'_x(x_0, y_0)(x - x_0) \\ + f'_y(x_0, y_0)(y - y_0) \quad (27)$$

where $f'_x(x_0, y_0)$ and $f'_y(x_0, y_0)$ are given by

$$f'_x(x_0, y_0) = \frac{-\gamma A_2 B}{\ln 2(x_0 + X)(\gamma(A_1(x_0 + X) + A_2) + (x_0 + X)(y_0 + Y)^{a_0/2})}$$

$$f'_y(x_0, y_0) = \frac{-\gamma a_0 B(A_1(x_0 + X) + A_2)}{\ln 4(y_0 + Y)(\gamma(A_1(x_0 + X) + A_2) + (x_0 + X)(y_0 + Y)^{a_0/2})} \quad (28)$$

We substitute $x_0 = 0$, $y_0 = 0$, $x = e^{-(K_3+K_4 v_{i,j})} - e^{-(K_3+K_4 \widehat{v_{i,j}})}$, $X = 1 + e^{-(K_3+K_4 \widehat{v_{i,j}})}$, $y = \|q_j - w_i\|^2 - \|\widehat{q_j} - w_i\|^2$, $Y = \|\widehat{q_j} - w_i\|^2 + H_j^2$ and $\gamma = \gamma_i$ into eq. (27), and can derive that

$$B\log_2\left(1 + \left(A_1 + \frac{A_2}{1+e^{-(K_3+K_4 v_{i,j})}}\right)\frac{\gamma_i}{(d_{i,j}^2)^{a_0/2}}\right)$$
$$\geq B\log_2\left(1 + \left(A_1 + \frac{A_2}{1+e^{-(K_3+K_4 \widehat{v_{i,j}})}}\right)\frac{\gamma_i}{(\widehat{d_{i,j}}^2)^{a_0/2}}\right)$$
$$- \frac{\gamma A_2 B}{\ln 2 X(\gamma(A_1 X + A_2) + X(Y)^{a_0/2})}x$$
$$- \frac{\gamma \kappa B(A_1 X + A_2)}{\ln 4 Y(\gamma(A_1 X + A_2) + X(Y)^{a_0/2})}y = r_{i,j}^{\text{lb}} \qquad (29)$$

where $r_{i,j}^{\text{lb}}$ is the lower bound for $r_{i,j}$. □

## APPENDIX D

*Proof of Theorem 4:* We define $f(x) = \frac{H_j}{\sqrt{x+X}}$. The second derivative of $f(x)$ is given by

$$f''(x) = \frac{3H_j}{4(x+X)^{5/2}} \geq 0 \qquad (30)$$

which means that $f(x)$ is convex with respect to $x$, and we can use the first-order Taylor expansion to approximate the convex function $f(x)$ at any $x_0$. Therefore, we can get

$$f(x) \geq f(x_0) + f'(x_0)(x - x_0) \qquad (31)$$

We substitute $x_0 = 0$, $x = \|q_j - w_i\|^2 - \|\widehat{q_j} - w_i\|^2$ and $X = \|\widehat{q_j} - w_i\|^2 + H_j^2$ into eq. (31), and can derive that

$$\frac{H_j}{\sqrt{\|q_j - w_i\|^2 + H_j^2}} \geq \frac{H_j}{\sqrt{\|\widehat{q_j} - w_i\|^2 + H_j^2}} - \frac{H_j}{2(\|\widehat{q_j} - w_i\|^2 + H_j^2)^{3/2}}x$$
$$= v_{i,j}^{\text{lb}} \qquad (32)$$

where $v_{i,j}^{\text{lb}}$ is the lower bound for $v_{i,j}$. □

## APPENDIX E

*Proof of Theorem 5:* We define $f(x,y) = x - \frac{y}{\sqrt{A_3+y^2}}$, the Hessian of $f(x,y)$ is

$$\nabla^2 f(x,y) = \begin{bmatrix} \frac{\partial^2 f(x,y)}{\partial x^2} & \frac{\partial^2 f(x,y)}{\partial x \partial y} \\ \frac{\partial^2 f(x,y)}{\partial y \partial x} & \frac{\partial^2 f(x,y)}{\partial y^2} \end{bmatrix} = \begin{bmatrix} 0 & 0 \\ 0 & \frac{3A_3 y}{(y^2+A_3)^{5/2}} \end{bmatrix} \qquad (33)$$

Then, the characteristic equation of $\nabla^2 f(x,y)$ is given by

$$|\lambda E - \nabla^2 f(x,y)| = \begin{vmatrix} \lambda & 0 \\ 0 & \lambda - \frac{3A_3 y}{(y^2+A_3)^{5/2}} \end{vmatrix}$$
$$= \lambda\left(\lambda - \frac{3A_3 y}{(y^2+A_3)^{5/2}}\right) = 0 \qquad (34)$$

and $\lambda_1 = 0$, $\lambda_2 = \frac{3A_3 y}{(y^2+A_3)^{5/2}} \geq 0$. We can prove that $\nabla^2 f(x,y)$ is a positive semidefinite matrix and $f(x,y)$ is a convex function. Therefore, the constraint of (15.2) is a convex constraint. □

## APPENDIX F

*Proof of Theorem 6*: To deal with non-convex constraint (19.3), we follow a similar approach as for solving problem (19). We substitute $x_0 = 0$, $y_0 = 0$, $x = e^{-(K_3+K_4 v_{i,j})} - e^{-(K_3+K_4 \widehat{v_{i,j}})}$, $X = 1 + e^{-(K_3+K_4 \widehat{v_{i,j}})}$, $y = H_j^2 - \widehat{H_j^2}$, $Y = \|q_j - w_i\|^2 + \widehat{H_j^2}$ and $\gamma = \gamma_i$ into eq. (27), and can derive that

$$r_{i,j} \geq \widetilde{r_{i,j}^{\text{lb}}} \triangleq \widehat{r_{i,j}} - \phi_{i,j}^X\left(e^{-(K_3+K_4 v_{i,j})} - e^{-(K_3+K_4 \widehat{v_{i,j}})}\right)$$
$$- \phi_{i,j}^Y\left(H_j^2 - \widehat{H_j^2}\right) \qquad (35)$$

where $\phi_{i,j}^X$ and $\phi_{i,j}^Y$ are given by

$$\phi_{i,j}^X = \frac{\gamma \kappa B(A_1 X + A_2)}{\ln 4 Y(\gamma(A_1 X + A_2) + X(Y)^{a_0/2})}$$
$$\phi_{i,j}^Y = \frac{\gamma A_2 B}{\ln 2 X(\gamma(A_1 X + A_2) + X(Y)^{a_0/2})} \qquad (36)$$